\documentclass[%
 aip,
 jap,% 
 amsmath,amssymb,
 reprint,%
]{revtex4-1}

\usepackage{graphicx}% Include figure files
\usepackage{dcolumn}% Align table columns on decimal point
\usepackage{bm}% bold math
\begin{document}

\preprint{AIP/123-QED}

\title{Feasibility, Accuracy and Performance of Contact Block Reduction method for multi-band simulations of ballistic quantum transport}% Force line breaks with \\

\author{Hoon Ryu}
\email{elec1020@gmail.com}
\affiliation{ 
Supercomputing Center, Korea Institute of Science and Technology Information, Daejeon 305-806, Republic of Korea. %\\This line break forced with \textbackslash\textbackslash
}%
\affiliation{
Network for Computational Nanotechnology, Purdue University, West Lafayette, Indiana 47907, USA.
}

\author{Hong-Hyun Park}%
\affiliation{
Network for Computational Nanotechnology, Purdue University, West Lafayette, Indiana 47907, USA.
}

\author{Mincheol Shin}
\affiliation{%
Department of Electrical Engineering, Korea Advanced Insistute of Science and Technology, Daejeon 305-701, Republic of Korea.
}%

\author{Dragica Vasileska}
\affiliation{%
Department of Electrical Engineering, Arizona State University, Tempe, Arizona 85287, USA.
}%

\author{Gerhard Klimeck}
\affiliation{
Network for Computational Nanotechnology, Purdue University, West Lafayette, Indiana 47907, USA.
}

\date{\today}% It is always \today, today,

\begin{abstract}
Numerical utilities of the Contact Block Reduction (CBR) method in evaluating the retarded Green's function, are discussed for 3-D multi-band open systems that are 
represented by the atomic tight-binding (TB) and continuum $k$$\cdot$$p$ (KP) band model. It is shown that the methodology to approximate solutions of open systems 
which has been already reported for the single-band effective mass model, cannot be directly used for atomic TB systems, since the use of a set of zincblende crystal grids 
makes the inter-coupling matrix be non-invertible. We derive and test an alternative with which the CBR method can be still practical in solving TB systems. This $multi$-$band$ 
$CBR$ method is validated by a proof of principles on small systems, and also shown to work excellent with the KP approach. Further detailed analysis on the accuracy, 
speed, and scalability on high performance computing clusters, is performed with respect to the reference results obtained by the state-of-the-art Recursive Green's Function 
and Wavefunction algorithm. This work shows that the CBR method could be particularly useful in calculating resonant tunneling features, but show a limited practicality in 
simulating field effect transistors (FETs) when the system is described with the atomic TB model. Coupled to the KP model, however, the utility of the CBR method can be 
extended to simulations of nanowire FETs.
\end{abstract}

%Use showkeys class option if keyword
\maketitle

\section{Introduction}

\subsection{Needs for multi-band approaches}

Semiconductor devices have been continuously downscaled ever since the invention of the first transistor \cite{MOORESLAW}, such that the size of the single building 
component of modern electronic devices has already reached to a few nanometers (nm). In such a $nanoscale$ regime, two conceptual changes are required in the 
device modeling methodology. One aspect is widely accepted where carriers must be treated as quantum mechanical rather than classical objects. The second change is the 
need to embrace the multi-band models which can describe atomic features of materials, reproducing experimentally verified bulk bandstuructures. While the single-band 
effective mass approximation (EMA) predicts bandstructures reasonably well near the conduction band minimum (CBM), the subband quantization loses accuracy if devices 
are in a sub-nm regime \cite{LUISIERPRB}. The EMA also fails to predict indirect gaps, inter-band coupling and non-parabolicity in bulk bandstructures \cite{KLIMECKIEEE}.

The nearest-neighbor empirical tight-binding (TB) and next nearest-neighbor $k$$\cdot$$p$ (KP) approach are most widely used band models of multiple bases 
\cite{KLIMECKIEEE, KDOTP}. The most sophisticated TB model uses a set of 10 localized orbital bases (s, s*, 3$\times$p, and 5$\times$d) on real atomic grids (20 with 
spin interactions), where the parameter set is fit to reproduce experimentally verified bandgaps, masses, non-parabolic dispersions, hydrostatic and biaxial strain behaviors 
of bulk materials using a global minimization procedure based on a genetic algorithm and analytical insights \cite{KLIMECKIEEE, KLIMECKVLSI, BOYKINPRB}. This  
${sp^3d^5s^*}$ TB approach can easily incorporate atomic effects such as surface roughness and random alloy compositions as the model is based on a set of atomic 
grids. These physical effects have been shown to be critical to the quantitative modeling of Resonance Tunneling Diodes (RTDs), quantum dots, disordered SiGe/Si 
quantum wells, and a single impurity device in Si bulk \cite{BOWEN, KHARCHEVALLEYSPLITTING, RAJIBPRL, LANSBERGNATURE}.

The KP approach typically uses four bases on a set of cubic grids with no spin interactions \cite{KDOTP}. While it still fails to predict the indirect gap of bulk dispersions 
since it assumes that all the subband minima are placed on the $\Gamma$ point, the credibility is better than the EMA since the KP model can still explain the inter-band 
physics of direct gap III-V devices, and valence band physics of indirect gap materials such as silicon (Si) \cite{KDOTPTRANSPORT, KDOTPOPTICS}.

\subsection{Contact Block Reduction method}

One of the important issues in modeling of nanoscale devices, is to solve the quantum transport problem with a consideration of real 3-D device geometries. Although
the Non-Equilibrium Green's Function (NEGF) and WaveFunction (WF) formalism have been widely used to simulate the carrier transport \cite{LUISIERPRB, KDOTPTRANSPORT, DATTANEGF, WAVEFUNCTION, KHARCHEIEDM}, 
the computational burden has been always a critical problem in solving 3-D open systems as the NEGF formalism needs to invert a system matrix of a degree-of-freedom 
(DOF) equal to the Hamiltonian matrix \cite{DATTANEGF}. The Recursive Green's Function (RGF) method saves the computing load by selectively targeting elements needed 
for the matrix inversion \cite{LAKE, RIVAS}. However, the cost can be still huge depending on the area of the transport-orthogonal plane (cross-section) and the length along 
the transport direction of target devices \cite{CAULEYJAP, MMLUYPRB}. The WF algorithm also saves the computing load if the transport is ballistic as it doesn't have to invert 
the system matrix and finding a few solutions of the linear system is enough to predict the transport behaviors. But, the load still depends on the size of the system matrix and 
the number of solution vectors (modes) needed to describe the carrier-injection from external leads \cite{LUISIERPRB, WAVEFUNCTION}. In fact, RGF and WF calculations 
for atomically resolved nanowire field effect transistors (FETs) have demonstrated the need to consume over 200,000 parallel cores on large supercomputing clusters \cite{CISE}.

Developed by Mamaluy \emph{et al.} \cite{MMLUYPRB, MMLUYSSTECH}, the Contact Block Reduction (CBR) method has received much attention due to the utility to save 
computing expense required to evaluate the retarded Green's function of 3-D open systems. The CBR method is thus expected to be a good candidate for transport simulations 
since the method doesn't have to solve the linear system yet reducing the computing load needed for matrix inversion \cite{MMLUYPRB}. The method indeed has been extensively 
used such that it successfully modeled electron quantum transport in experimentally realized Si FinFETs \cite{CBRTED01}, and predicted optimal design points and process 
variations in design of 10-nm Si FinFETs \cite{CBRTED02, CBRTED03}. However, all the successful applications for 3-D systems so far, have been demonstrated only for the 
systems represented by the EMA.

\subsection{Goals of this work}

While the use of multi-band approaches can increase the accuracy of simulation results, it requires more computing load as a DOF of the Hamiltonian matrix is directly proportional 
to the number of bases required to represent a single atomic (or grid) spot in the device geometry. To suggest a solution to this \emph{trade-off} issue, we examine the numerical 
utilities of the CBR method in multi-band ballistic quantum transport simulations, focusing on multi-band 3-D systems represented by either of the TB or KP band model. 

The objective of this work is to provide detail answers to the following questions through simulations of small two-contact ballistic systems focusing on a proof of principles: (1) Can 
the original CBR method be extended to simulate ballistic quantum transport of multi-band systems? (2) If the answer to the question (1) is $``yes''$, what is the condition under which 
the multi-band CBR method becomes particularly useful?, and (3) How is the numerical practicality of the multi-band CBR method compared to the RGF and WF algorithms, in terms 
of the accuracy, speed and scalability on High Performance Computing (HPC) clusters? 

\section{Methodology}

In real transport problems, a device needs to be coupled with external contacts that allow the carrier-in-and-out flow. With the NEGF formalism, this can be done by creating 
an open system that is described with a non-Hermitian system matrix \cite{DATTANEGF}. Representing this system matrix as a function of energy, we compute the transmission 
coefficient and density of states, to predict the current flow and charge profile in non-equilibrium. This energy-dependent system matrix is called the retarded Green's function 
$G^R$ for an open system (Eq. (1)).

\begin{equation}
G^{R}(E)=[(E+i\eta)I-H^o-\Sigma(E)]^{-1}, \indent \eta \rightarrow 0^{+} 
\end{equation}

\noindent 
where ${H^o}$ is  is the Hamiltonian representing the device and $\Sigma$ is the self-energy term that couples the device to external leads. As already mentioned in the previous 
section, the evaluation of $G^R$ is quite computationally expensive since it involves intensive matrix inversions. The CBR method, however, reduces matrix inversions with the 
mathematical process based on the Dyson equation. We start the discussion revisiting the CBR method that has been so far utilized for EMA systems.

\subsection{Revisit: CBR with EMA}

The CBR method starts decomposing the device domain into two regions: (1) the boundary region $c$ that couples with the contacts, and (2) the inner region $d$ that doesn't 
couple to the contacts. As the self-energy term $\Sigma$ is non-zero only in the boundary region, ${H^o}$ and $\Sigma$ are decomposed as shown in Eq. (2), where subscripts 
($c$, $d$) denote above-mentioned regions, respectively. 

\begin{equation}
H=
\begin{bmatrix}
H_c^o & H_{cd}^o \\
H_{dc}^o & H_d^o
\end{bmatrix}
,\indent\Sigma=
\begin{bmatrix}
\Sigma_c & 0_{cd} \\
0_{dc} & 0_d\
\end{bmatrix}
\end{equation} 

Then, ${G^R}$ can be evaluated with the Dyson equation defined in Eq. (3) and Eq. (4), where ${\Sigma^{x}}$ and ${G^{x}}$ are conditioned with a Hermitian matrix $X$ to minimize 
matrix inversions by solving the eigenvalue problem (Eqs. (5)).

\begin{equation}
A_c^{-1}=(I_c-G_c^x\Sigma_c^x)^{-1}
\end{equation}

\begin{eqnarray}
G^R(E) & = & (I-\Sigma^xG^x)^{-1}G^x \\
& = &
\begin{bmatrix}
A_c^{-1} & 0_{cd} \\
-G_{dc}^x\Sigma_c^xA_c^{-1} & I_d
\end{bmatrix}
\begin{bmatrix}
G_c^x & G_{cd}^x \\
G_{dc}^x & G_d^x 
\end{bmatrix} \nonumber \\
\nonumber \\ \nonumber \\
X & = & 
\begin{bmatrix}
x_c & 0_{cd} \\
0_{dc} & 0_d
\end{bmatrix}
,\indent\Sigma^x=\Sigma-X, \\
G^x & = &[EI-(H^o+X)]^{-1} \nonumber \\
& = &
\begin{bmatrix}
G_c^x & G_{cd}^x \\
G_{dc}^x & G_d^x 
\end{bmatrix} 
= \sum_\alpha\frac{|\Psi_\alpha\rangle\langle\Psi_\alpha|}{E-\epsilon_\alpha+i\eta} \nonumber 
\end{eqnarray}

\noindent
where ${\epsilon_\alpha}$ and ${\Psi_\alpha}$ are the ${\alpha^{th}}$ eigenvalue and eigenvector of the modified Hamiltonian (${H^o}$+$X$). Here, we note that the matrix inversion 
is performed only to evaluate the boundary block $A_c$ (contact-block) for one time while the RGF needs to perform the block-inversion many times depending on the device channel 
length. The computing load for matrix inversion is thus significantly reduced, and the method is also free from solving a linear-system problem. Instead, the major numerical issue now 
becomes a normal eigenvalue problem for a Hermitian matrix (${H^o}$+$X$). For the numerical practicality, it is thus critical to reduce a number of required eigenvalues, and for EMA 
Hamiltonian matrices, a huge reduction in the number of required eigenvalues can be achieved via a smart choice of the \emph{prescription matrix} $X$. 

\begin{figure}[t]
\centering
\includegraphics[width=\columnwidth]{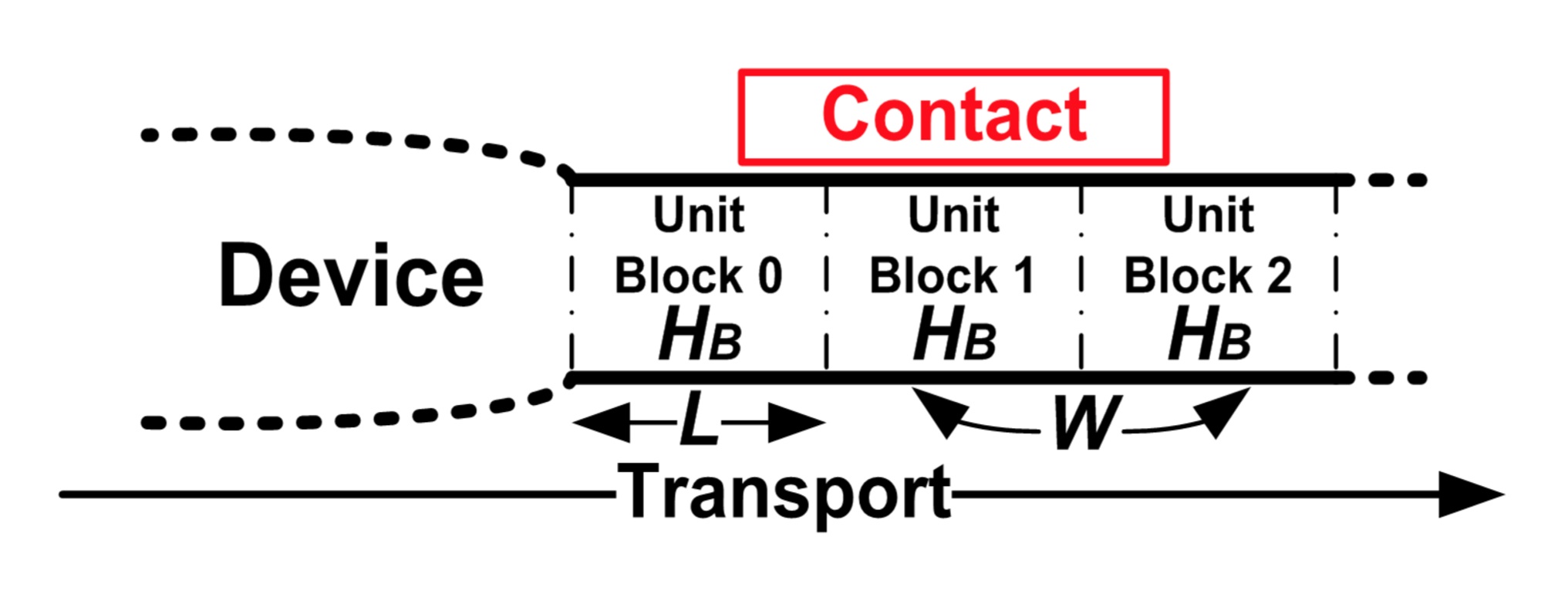}
\caption{
	Schematic of the semi-infinite contact to illustrate the treatment of the external contact that is normally 
	assumed to be an infinite chain of the slab on the device boundary (the outmost slab in the device domain). 
	      }
\end{figure}

To find the matrix $X$ and see if it can be extended to multi-band systems, we first need to understand how to couple external contacts to the device. Fig. 1 illustrates the common 
approach which treats the contact as a semi-infinite nanowire of a finite cross-section. Here, ${H_B}$ is a block matrix that represents the unit-slab along the transport direction, and 
$W$ is another block matrix which represent the inter-slab coupling. The eigenfunction of the plane wave at the ${m^{th}}$ mode in the ${n^{th}}$ slab, $\Psi_{(n,m)}$ should then 
obey the Schr\"odinger equation and the Bloch condition (Eqs. (6)).

\begin{eqnarray}
(EI-H_B)\Psi_{(n,m)} = W^+\Psi_{(n-1,m)} + W\Psi_{(n+1,m)}, \nonumber \\
\Psi_{(n+1,m)} = exp(ik_mL)\Psi_{(n,m)} \indent(1\leq m \leq M) 
\end{eqnarray}

\noindent where ${k_m}$ is the plane-wave vector at the ${m^{th}}$ mode, $L$ is the length of a slab along the transport direction, and M is the maximum number of plane-wave modes 
that can exist in a single slab and is equal to the DOF of ${H_B}$. Then, the surface Green's function ${G_{surf}}$ and self-energy term $\Sigma$ can be evaluated by converting Eqs. (6) 
to the generalized eigenvalue problem for a complex and non-Hermitian matrix \cite{RIVAS}. The solution for ${G_{surf}}$ and $\Sigma$ are provided in Eqs. (7), where $K$ and $\Lambda$ 
are shown in Eqs. (8).

\begin{eqnarray}
G_{surf} & = & K[K^{-1}(H_B-EI)K+K^{-1}W^{+}K\Lambda]^{-1}K^{-1}, \nonumber \\
\Sigma & = & W^+G_{surf} W
\end{eqnarray}

\begin{eqnarray}
K & = & [\Psi_{(0,1)}\text{ }\Psi_{(0,2)}\text{ }\ldots\text{ }\Psi_{(0,M)}], \nonumber \\
\Lambda & = & diag[exp(ik_1L)\text{ }exp(ik_2L)\ldots\text{ }exp(ik_ML)]
\end{eqnarray}

In systems described by the nearest-neighbor EMA, each slab becomes a layer of common cubic grids such that each grid on one layer is coupled to the same grid on the nearest layer. 
The inter-slab coupling matrix $W$ thus becomes a ${scaled\text{ }identity\text{ }matrix}$, with which the general solution for ${G_{surf}}$ and $\Sigma$ in Eqs. (7) can be simplified 
using a process described in Eq. (9) and Eq. (10). We note that previous literatures have shown only the simplified solution for ${G_{surf}}$ and $\Sigma$ \cite{MMLUYPRB, MMLUYSSTECH}.

\begin{eqnarray}
G_{surf} & = & K[K^{-1}(H_B-EI)K+K^{-1}W^+K\Lambda]^{-1}K^{-1} \nonumber \\
& = & K[K^{-1}(H_{B}-EI)K+W^+\Lambda]^{-1}K^{-1} \nonumber \\
& = & K[-K^{-1}(W^+K\Lambda+WK\Lambda^{-1})+W^+\Lambda]^{-1}K^{-1} \nonumber \\
&& [\because (EI-H_{B})K=W^+K\Lambda+WK\Lambda^{-1}] \nonumber \\
& = &-K[W\Lambda^{-1}]^{-1}K^{-1} = -K\Lambda W^{-1}K^{-1} \\
\nonumber \\ 
\Sigma & = & W^+G_{surf}W \nonumber \\
& = & W^+(-K\Lambda W^{-1}K^{-1})W \nonumber \\
& = & -W^+K\Lambda K^{-1}\text{   }(\because W^+ = W) \nonumber \\  
& = & -WK\Lambda K^{-1}
\end{eqnarray}

The original CBR method coupled to the EMA prescribes the Hermitian matrix $X$ as $-W$ or its Hermitian component (if $W$ is complex). The new self-energy term $\Sigma_x$ in 
Eqs. (5) then becomes (Eq. (11)):

\begin{eqnarray}
\Sigma_x & = & \Sigma - X = \Sigma + W \nonumber \\
& = & -WK\Lambda K^{-1} + W \nonumber \\
& = & -WK(\Lambda-I)K^{-1}
\end{eqnarray}

\noindent where the matrix ($\Lambda-I$) becomes zero at $\Gamma$ point, where EMA subband minima are always placed on. The resulting new Hamitonian (${H^o}$-$W$), becomes 
the Hamiltonian with the generalized $Von$-$Neumann$ boundary condition at contact boundaries. The spectra of the matrix (${H^o}$-$W$), therefore become approximate solutions of 
the open boundary problem, and the retarded Green's function $G^R(E)$ in Eq. (4) can be thus $approximated$ with an incomplete set of energy spectra of the Hermitian matrix near 
subband minima \cite{MMLUYPRB, MMLUYJAP}.

\subsection{CBR with multi-band models}

Regardless of the band model, the $G^R(E)$ in Eq. (4) can be accurately calculated with a complete set of spectra since it then becomes the Dyson equation (Eq. (3)) itself. The important 
question here is then whether we can make the CBR method be still numerically practical for multi-band systems such that the transport can be simulated with a narrow energy spectrum. 
To study this issue, we focus on the inter-slab coupling matrix $W$ of multi-band systems. A toy Si device that consists of two slabs along the [100] direction, is used as an example for our 
discussion.  

Fig. 2 shows the device geometry and corresponding Hamiltonian matrix built with the EMA, KP and TB model, respectively. Here, we note that the simplifying process in Eq. (9) and Eq. 
(10) is not strictly correct if the inter-slab coupling matrix $W$ is not an identity matrix, since, for any square matrix $K$ and $W$, $K^{-1}WK$ cannot be simplified to $W$ if $W$ is neither 
an identity matrix nor a scaled identity matrix. When a system is represented with KP model, a single slab is still a layer of common cubic grids as the KP approach also uses a set of cubic 
grids. But, the non-zero coupling is extended up to next-nearest neighbors such that the inter-slab coupling matrix $W$ is no more an identity matrix. The simplified solution for $G_{surf}$ 
and $\Sigma$, however, can be still used to $approximate$ the general solutions in Eqs. (7) since the coupling matrix $W$ is $diagonally\text{ }dominant$ and invertible. But, the situation 
becomes tricky for TB systems that are represented on a set of real zincblende (ZB) grids. 

\begin{figure}[t]
\centering
\includegraphics[width=\columnwidth]{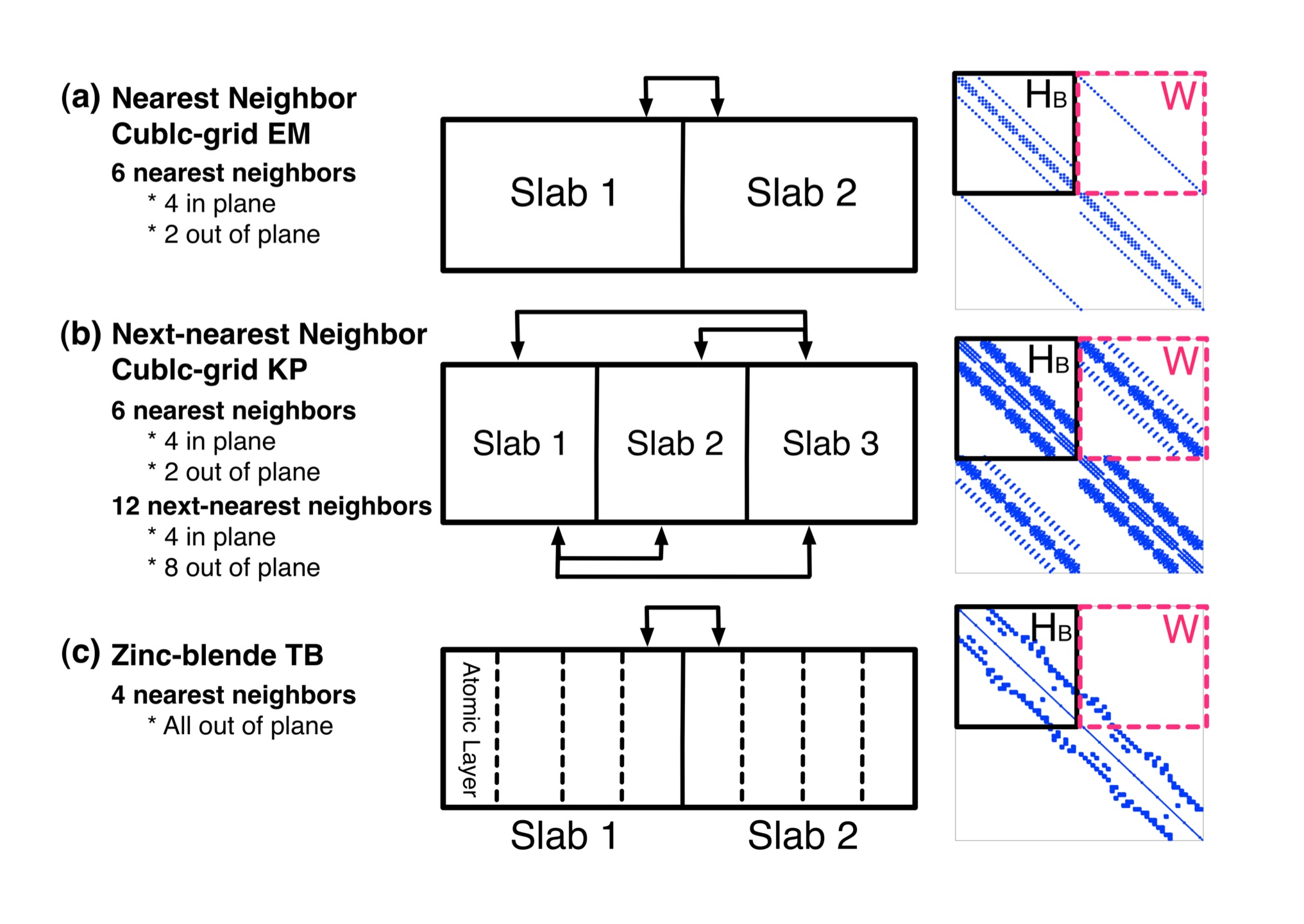}
\caption{
	Illustration of the geometry and the Hamiltonian matrix built for the (a) EMA, (b) KP, and (c) TB toy nanowire. Arrows represent 
	the inter-slab coupling. The simplifying process in Eq. (9) and Eq. (10) are not strictly accurate for multi-band models since the 
	inter-slab coupling matrix is neither an identity matrix nor a scaled identity matrix. Especially, the coupling matrix becomes singular 
	in TB model, which indicates that the simplified solution for $G_{surf}$ and $\Sigma$ are even mathematically invalid. 
	}
\end{figure}

In the ZB crystal structure, a Si unit-slab has a total of four unique atomic layers along the [100] direction. Because the TB approach assumes the nearest-neighbor coupling, only the last 
layer in one slab is coupled to the first layer in the nearest slab while all the other coupling blocks among layers in different slabs become zero-matrices. As described in Fig. 2, this makes 
the inter-slab coupling matrix $W$ be $singular$ such that matrix inversions become impossible. The simplified solution for $G_{surf}$ and $\Sigma$ in Eq. (9) and Eq. (10) are therefore 
mathematically invalid, and they cannot be even used to approximate the full solution (Eqs. (7)). A new prescription for $X$ is thus needed to make the CBR method be still practical for 
ZB-TB systems, and we propose an alternative in Eq. (12).

\begin{equation}
X=\frac{\Sigma(\epsilon_{edge}) + \Sigma^+(\epsilon_{edge})}{2} 
\end{equation}

\noindent where ${\epsilon_{edge}}$ is the energetic position of the CBM (valence band maximum (VBM)) of the bandstructure of the semi-infinite contact. 

If only a few subbands near the CBM (or VBM) of the contact bandstructure are enough to describe the external contact, the prescription suggested in Eq. (12) works quite well as $X$ 
is the Hermitian part of the self-energy term, such that (${H^o+X}$) approximates the open system near the edge of the contact bandstructure, The approximation, however, becomes 
less accurate if more subbands in higher energy (in lower energy for valence band) are involved to the open boundaries. Away from the band edge, subband placement becomes 
denser and inter-subband coupling becomes stronger. The prescription $X$ in Eq. (12) then would not be a good choice as it only approximate the open boundary solution near 
band edges, and the CBR method thus needs more eigenspectrums to solve open boundary transport problems. So, for example, the multi-band CBR method would not be numerically 
practical to simulate FETs at a high source-drain bias, since a broad energy spectrum is then needed to get an accurate solution.

Before closing this section, we note that, if the inter-slab coupling matrix $W$ is either an identity matrix or a scaled identity matrix, the prescription matrix $X$ in Eq. (12) becomes 
$identical$ to the one utilized to simulate 3-D systems in the previous literatures \cite{CBRTED01, CBRTED02, CBRTED03}, where (${H^o+X}$) approximates the open system well 
near $every$ $subband$ $minima$ if the system is represented by the EMA \cite{MMLUYSSTECH, MMLUYPRB, MMLUYJAP}. Once $G_{surf}$ and $\Sigma$ are determined from 
the prescription matrix $X$, evaluation of the transmission coefficient (TR) and the density of states (DOS) can be easily done \cite{DATTANEGF, MMLUYSSTECH, MMLUYPRB, MMLUYJAP}. 
Further detailed mathematics regarding derivation of TR and DOS will not be thus discussed here.

\section{Results and Discussions}

The results are discussed in two subsections. First, we validate the CBR method for multi-band systems with the new prescription for $X$ in Eq. (12). Focusing on a proof of principles, 
we compute the TR and DOS profiles for a toy TB and KP system, compare the result to the references obtained with the RGF algorithm, and suggest the device category where the 
multi-band CBR method could be particularly practical. Second, we examine the numerical practicality of the multi-band CBR method by computing TR and DOS profiles of a resonant 
tunneling device and a nanowire FET. The accuracy, the speed of calculations in a serial mode, and the scalability on HPC clusters, are compared to those obtained with the RGF and 
WF algorithm. We assume a two-contact ballistic transport for all the numerical problems.

\subsection{Validation of multi-band CBR method}

To validate the multi-band CBR method that has been discussed in the previous section, we consider two multi-band toy Si systems represented by the 10-band ${sp^3d^5s^*}$ TB 
and 3-band KP approach. Here, we intentionally choose extremely small systems to calculate a complete set of energy spectra of the Hamiltonian, with which the CBR method should 
produce results identical to the ones obtained by the RGF algorithm. For the TB system, the electron-transport is simulated while we calculate the hole-transport for the KP system due 
to a limitation of the KP approach in representing the Si material \cite{NOTE01}. 

\emph{TB System}: Fig. 3 illustrates the TR and DOS profile calculated for the TB Si toy device which consists of (2$\times$2$\times$2) (100) unit-cells ($\sim$1.1(nm)). The device 
involves a complex Hermitian Hamiltonian matrix of 640 DOF, and electrons are assumed to transport along the [100] direction. The TR and DOS profiles are calculated using the 
CBR method for a total of three cases - with 6, 60 and full (640) energy spectra that correspond to 1$\%$, 10$\%$, and 100$\%$ of the Hamiltonian DOF, respectively. The transport 
happens at the energy above 2.32(eV) which is the CBM of the contact bandstructure. We note that this energetic position is higher than the Si bulk CBM (1.13(eV)), due to the 
structural confinement stemming from the finite cross-section of the nanowire device \cite{NOTE02}.

\begin{figure}[t]
\centering
\includegraphics[width=\columnwidth]{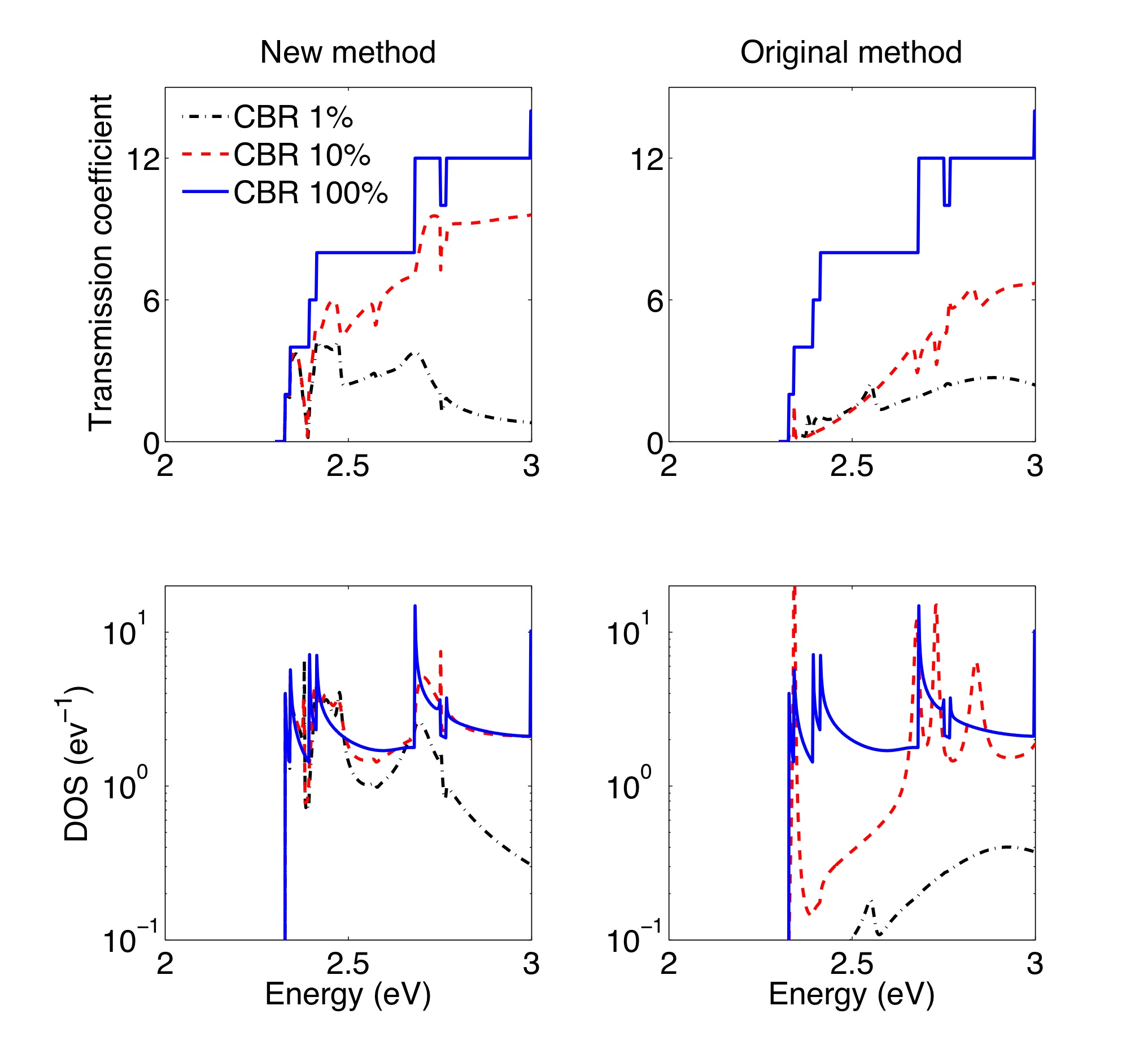}
\caption{
	(Electron-transport in a toy Si TB system) TR and DOS profiles calculated by the CBR method: Results with a prescription 
	suggested in Eq. (12) (New method), and an old prescription suggested for the EMA (Original method). Note that, with the 
	old prescription, using more energy spectra does not improve the accuracy of the CBR solution.
	}
\end{figure}

With the new prescription matrix $X$ suggested in Eq. (12), the TR and DOS profile obtained by the CBR method become closer to the reference result as more spectrums are used, 
and eventually reproduce the reference result with a full set of spectrums, as shown in the left column of Fig. 3. Here, the CBR result turns out to be quite accurate near the CBM even 
with 1$\%$ of the total spectrums, indicating that the TB-CBR method could be a practical approach if most of the carriers are injected from the first one or two subbands of the contact 
bandstructure. This condition can be satisfied when (1) only the first one or two subbands in the contact bandstructure are occupied with electrons, and (2) the energy difference between
the source and drain contact Fermi-level (the source-drain Fermi-window) becomes extremely narrow. So, the simulation of FETs at a high source-drain bias would not be an appropriate 
target of the TB-CBR simulations since the source-drain Fermi-window may include many subbands, and many spectra may be thus needed for accurate solutions \cite{NOTE03}. Instead, 
we propose that RTDs could be one of device categories for which the TB-CBR method is particularly practical, since the Fermi-window for transport becomes extremely small in RTDs in 
some cases \cite{LINDPRB}.

\begin{figure}[t]
\centering
\includegraphics[width=\columnwidth]{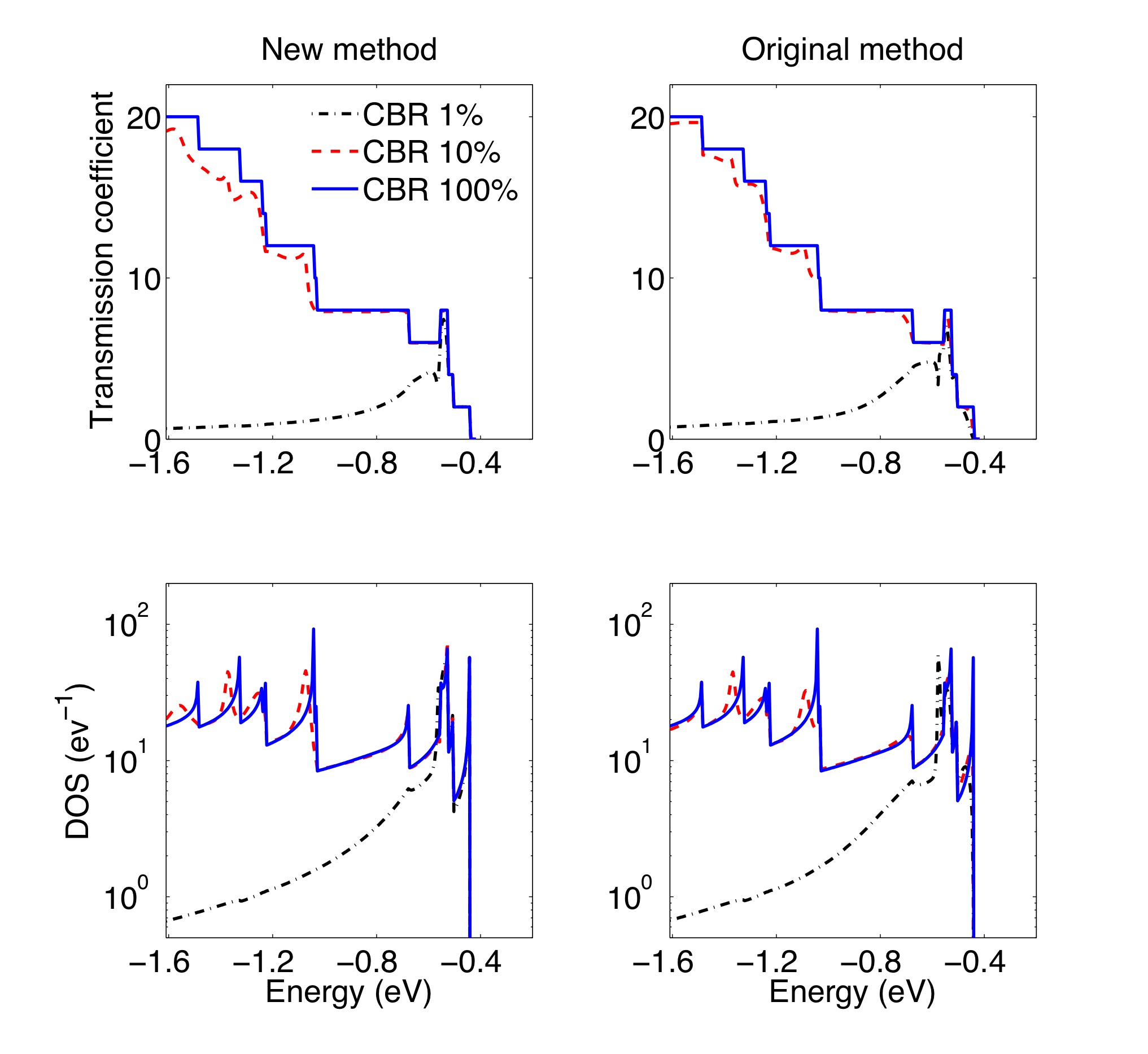}
\caption{
	(Hole-transport in a toy Si KP system) TR and DOS profiles calculated by the CBR method: Results with a prescription 
	suggested in Eq. (12) (New method), and an old prescription suggested for the EMA (Original method). Note that the 
	accuracy of the CBR solution is similar with both the new and old prescription.
	}
\end{figure}

The same calculation is performed again but using the old prescription $X$ suggested for the EMA, and corresponding TR and DOS profiles are shown in the right column of Fig. 3. The 
CBR method still reproduce the reference result with a full set of energy spectra since the Dyson equation (Eq. (4)) should always work for any $X$'s. The accuracy of the results near the 
CBM, however, turns out to be worse than the one with the new prescription. The results furthermore reveal that the accuracy with 10$\%$ of the total spectra does not necessarily becomes 
better than the one with 1$\%$, indicating that the old prescription for $X$ cannot even approximate the solution near the CBM of open TB systems.

\emph{KP System}: The TR and DOS profile of the KP Si 2.0(nm) (100) cube, are depicted in Fig. 4. The structure is discretized with a 0.2(nm) grid and involves a complex Hermitian Hamiltonian 
of 3,000 DOF. Here, the DOF of the real-space KP Hamiltonian can be effectively reduced with the $mode$-$space$ approach \cite{KDOTPTRANSPORT}. The effective DOF of the Hamiltonian 
therefore becomes 500, where we consider 50 modes per each slab along the transport direction. Again, we note that the VBM of the contact bandstructure is placed at -0.4(eV), and lower in 
energy than the VBM of Si bulk (0(eV)) due to the confinement created by the finite cross-section. 

We claim that the CBR method works quite well for the KP system, since the TR and DOS profiles not only become closer to the reference results as more of the energy spectrums are used, 
but also exhibit excellent accuracy near the VBM of the contact bandstructure as shown in Fig. 4. We, however, observe a remarkable feature that is not found in the CBR method coupled to 
TB systems: The KP-CBR method shows a good accuracy with both the old and new prescription matrix $X$, which supports that the simplified solution for ${G_{surf}}$ and $\Sigma$ (Eq. (9) 
and Eq. (10)) are still useful to approximate the full solution (Eqs. (7)) as discussed in the previous section. We also claim that the utility of the KP-CBR method could be extended to nanowire 
FETs because the mode-space approach reduces the DOF of the Hamiltonian such that we save more computing cost needed to calculate energy spectra. In the next subsection, we will come 
back to this issue again.

\begin{figure}[t]
\centering
\includegraphics[width=\columnwidth]{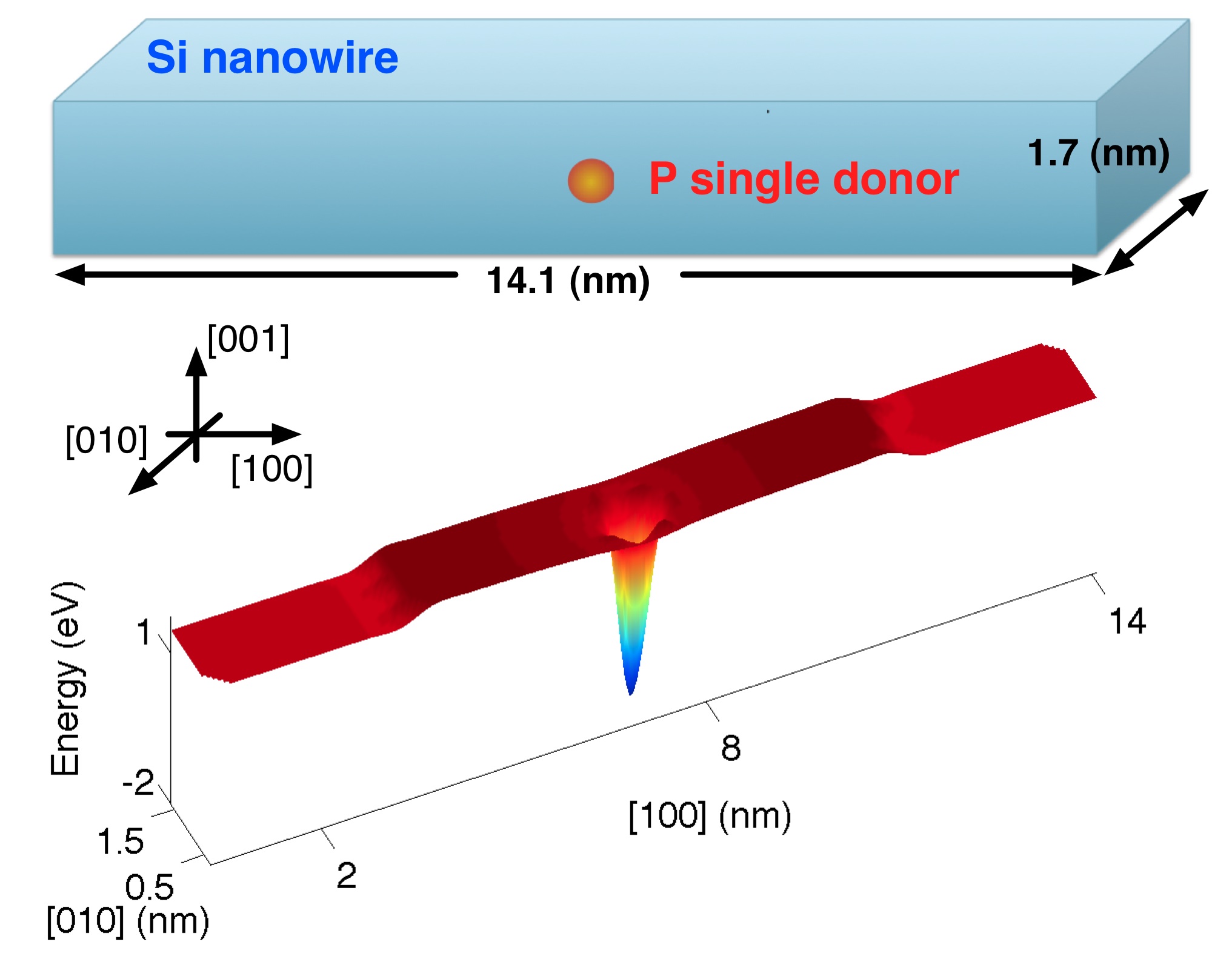}
\caption{
	Illustration of the geometry and potential profile of a Si:P RTD that is used as the example to examine the utility for the TB-CBR method.
	For the potential profile, 1.13(eV) is used as a reference value representing the Si Bulk CBM. The single donor coulombic potential that
	has been calibrated by Rahman \emph{et al.} (Ref. [9]), with respect to the Si bulk, is superposed to the channel potential profile to consider
	the sharp structural confinement stemming from the single donor. 
	}
\end{figure}

\subsection{Practicality of multi-band CBR method}

In this subsection, we provide a detailed analysis of the numerical utility of the multi-band CBR method in terms of the accuracy and speed. Based on discussions in the previous subsection 
with a focus on a proof of principles on small systems, a RTD is considered as a simulation example of TB systems, while a nanowire FET is again used as an example of KP systems to discuss 
the numerical practicality of the method. The TR and DOS profiles obtained by the RGF and WF algorithm are used as reference results. We note that the WF case is added in this subsection 
to provide a complete and competitive analysis on the speed and scalability on HPC clusters.

\begin{figure*}[t]
\centering
\includegraphics[width=\textwidth]{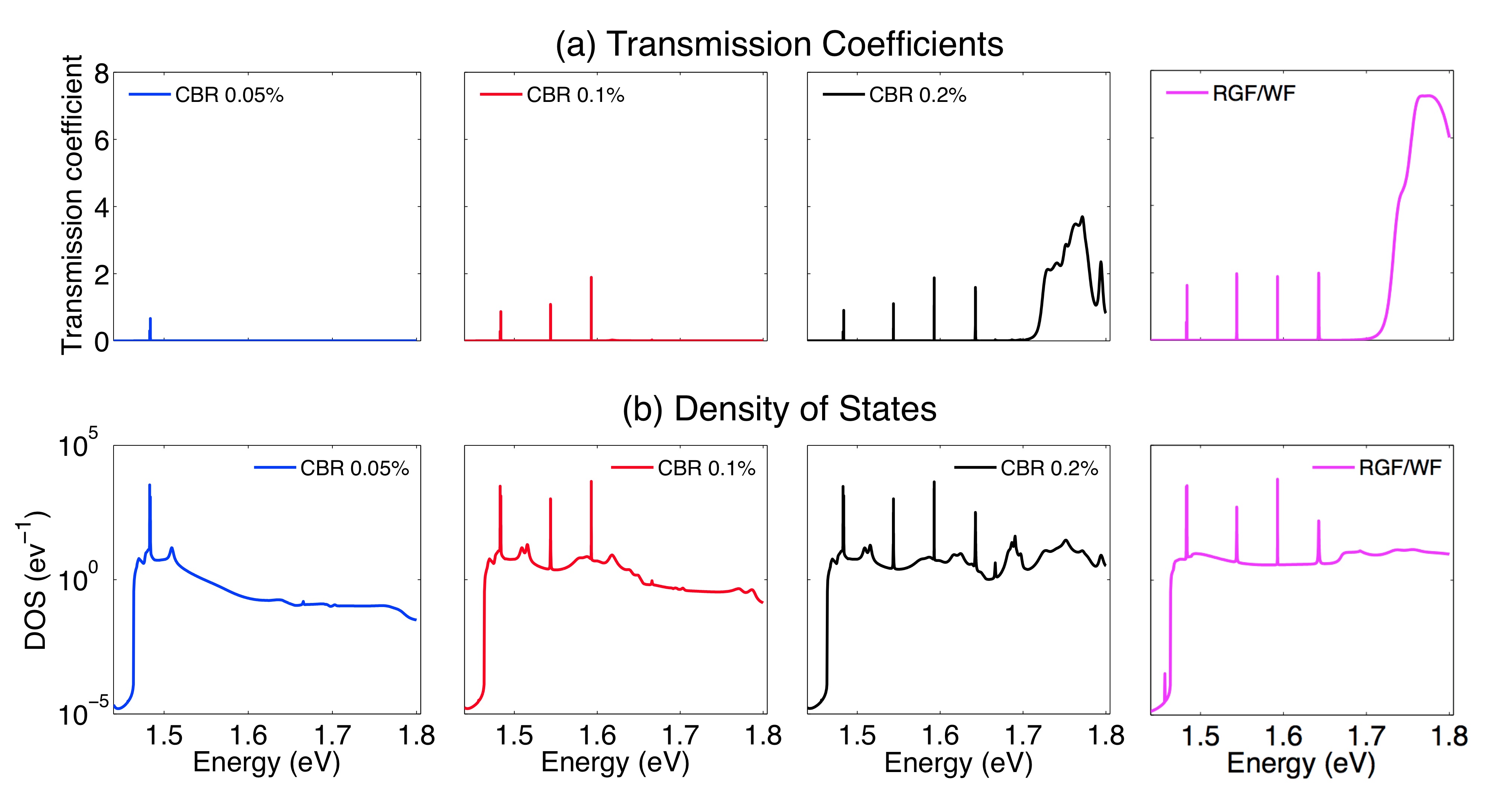}
\caption{
	(Electron-transport in a Si:P RTD) (a) TR and (b) DOS profiles calculated by the CBR method. 
	Note that all the resonances in the range of energy are captured even with just 40 spectra that
	corresponds to just 0.2$\%$ of the DOF of the TB Hamiltonian.
	}
\end{figure*}

\emph{TB System}: A single phosphorous donor in host Si material (Si:P) creates a 3-D structural confinement around itself. Such Si:P $quantum$ $dots$ have gained scientific interest due to 
their potential utility for qubit-based logic applications \cite{LLOYDQUBIT}. Especially, the Stark effect in Si:P quantum dots is one of the important physical problems, and was quantitatively 
explained by previous TB studies \cite{RAJIBPRL, LANSBERGNATURE}. The electron-transport in such Si:P systems should be therefore another important problem that needs to be studied.

\begin{figure}[t]
\centering
\includegraphics[width=\columnwidth]{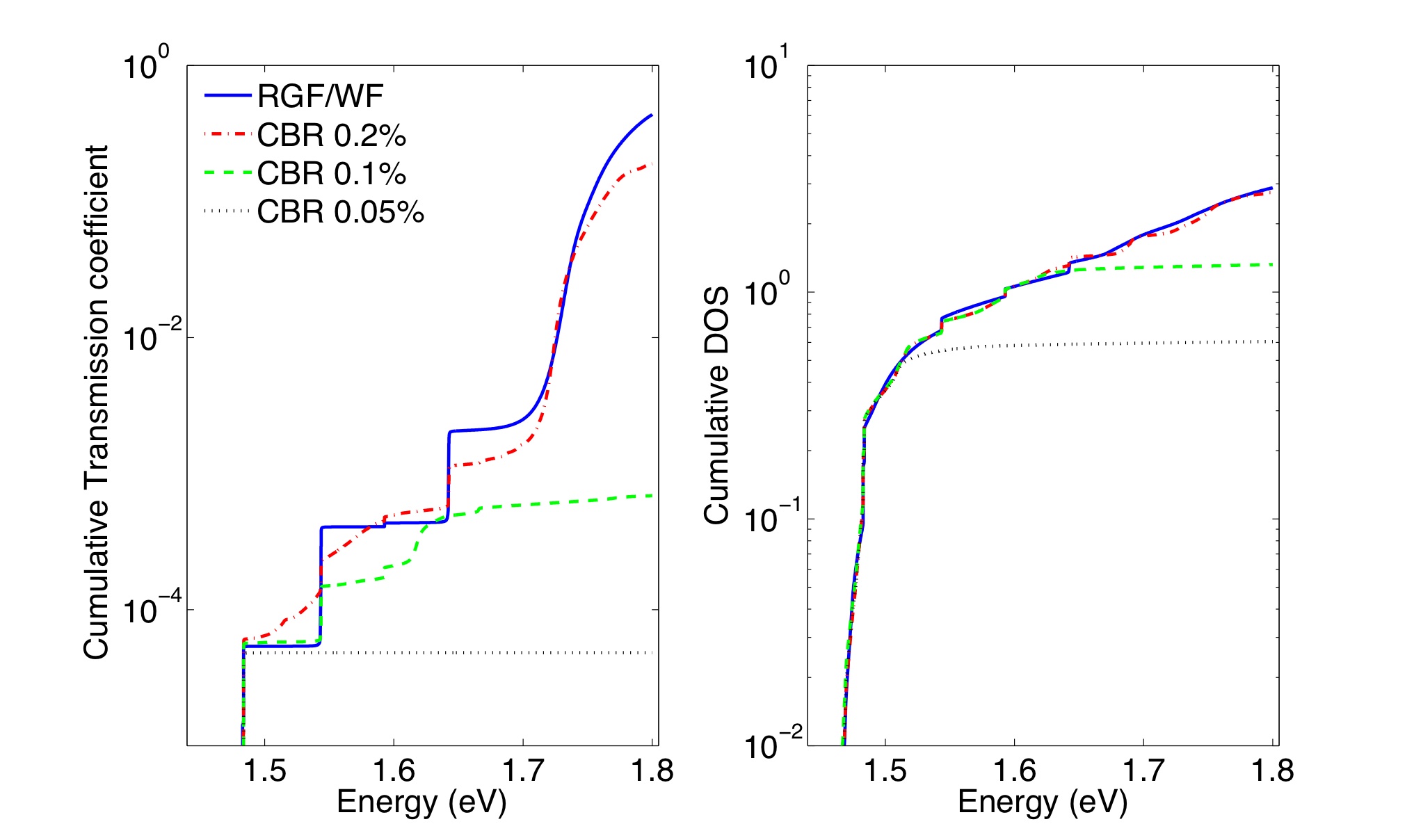}
\caption{
	(Electron-transport in a Si:P RTD) TR and DOS profiles integrated over energy. The cumulative profiles of TR and DOS 
	effectively indicates the accuracy of the current and charge profiles. The cumulative DOS is especially important as it is
	directly coupled to charge profiles that are needed for charge-potential self-consistent simulations.
	}
\end{figure}

The geometry of the example Si:P device is illustrated in Fig. 5. Here, we consider a [100] Si nanowire that is 14.0(nm) long and has a 1.7(nm) rectangular cross-section. The first and last 3.0(nm) 
along the transport direction, are considered as densely N-type doped source-drain region assuming a 0.25(eV) band-offset in equilibrium \cite{NOTE04}. Then, a single phosphorous atom is 
placed at the channel center with a superposition of the impurity coulombic potential that has been calibrated for a single donor in Si bulk by by Rahman \emph{et al.} \cite{RAJIBPRL}. The 
electronic structure has a total of 1872 atoms and involves a complex Hamiltonian matrix of 18,720 DOF.

Fig. 6 shows the TR and DOS profiles in four cases, where the first three cases are the CBR results with 10, 20 and 40 spectra that correspond to 0.05$\%$, 0.1$\%$, and 0.2$\%$ of the Hamiltonian 
DOF, and the last one is used as  a reference. Due to the donor coulombic potential, the channel forms a double-barrier system such that the electron transport should experience a resonance 
tunneling. As shown in Fig. 5, the CBR method produces a nice approximation of the reference result such that the first resonance is observed with just 10 energy spectra. It also turns out that 40 
spectra are enough to capture all the resonances that show up in the range of energy of interest. 

The accuracy of the solutions approximated by the CBR method, is examined in a more quantitative manner by $integrating$ the TR and DOS profile over energy. Fig. 7 illustrates this $cumulative$ 
TR (CTR) and DOS (CDOS) profile, which are $conceptually$ equivalent to the current and charge profile, respectively. In spite of a slight deviation in absolute values, the CTR profiles still confirm 
that the CBR method captures resonances quite precisely such that the energetic positions where the TR sharply increases, are almost on top of the reference result. The CDOS profile exhibits 
much better accuracy such that the result with 40 spectra almost reproduces the reference result even in terms of absolute values. We claim that the accuracy in the CDOS profile is particularly 
critical, since it is directly connected to charge profiles that are essential for charge-potential self-consistent simulations.

\begin{figure}[t]
\centering
\includegraphics[width=\columnwidth]{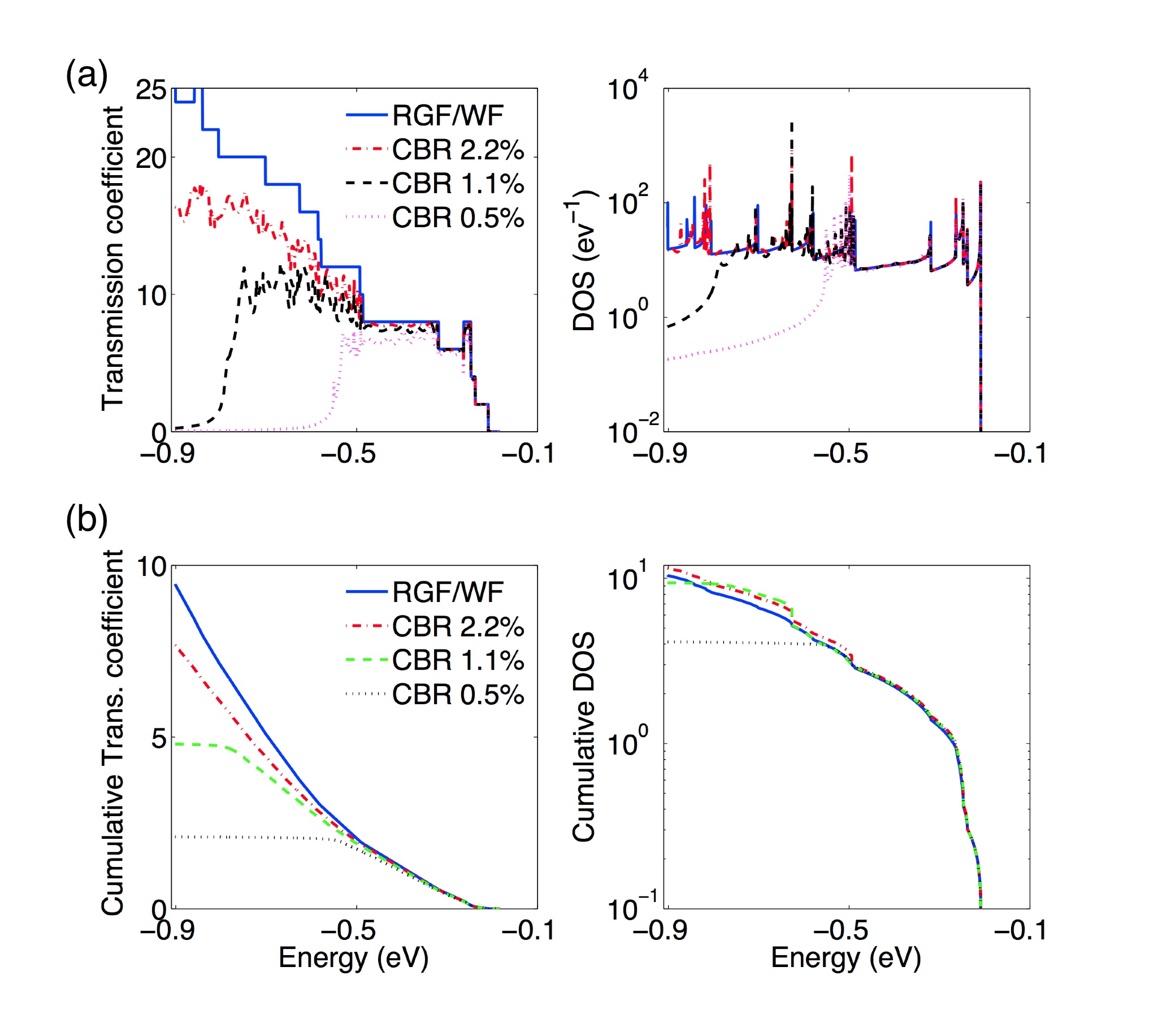}
\caption{
	(Hole-transport in a Si nanowire) (a) TR and DOS profiles, and (b) corresponding cumulative profiles. 
	The KP-CBR solutions exhibit excellent accuracy such that 200 spectra, which corresponds to just 2.2$\%$,
	turn out to be enough to almost reproduce the reference solutions in the entire range of energy of interest 
	(0.8(eV) beyond the VBM of the wire bandsturcutrue).
	}
\end{figure}

\emph{KP System}: Si nanowire FETs obtained through top-down etching or bottom-up growth have attracted attention due to their enhanced electrostatic control over the channel, and thus 
become an important target of various modeling works \cite{KDOTPTRANSPORT, PHYTOS}. For KP systems, the CBR method could become a practical approach to solve transport behaviors 
of FET devices since the computing load for solving eigenvalue problems can be reduced with the mode-space approach. 

A [100] Si nanowire FET of a 15.0(nm) long channel and a 3.0(nm) rectangular cross-section, is therefore considered as a simulation example to test the performance of the KP-CBR method. 
The hole-transport is simulated with the 3-band KP approach, where the simulation domain is discretized with a set of 0.2(nm) mesh cubic grids and involves a real-space Hamiltonian matrix 
of 50,625 DOF. As the device has a total of 75 slabs along the transport direction, the mode-space Hamiltonian has 9,000 DOF with a consideration of 120 modes per slab. It has been reported 
that the wire bandstructure obtained with 120 modes per slab, becomes quite close to the full solution for a cross-section smaller than 5.0$\times$5.0(nm$^2$) \cite{KDOTPTRANSPORT}. The 
wire is assumed to be purely $homogeneous$ such that neither the doping nor band-offset are considered.

To see if the CBR method can be reasonably practical in simulating the hole-transport at a relatively large source-drain bias, we plan to cover the energy range at least larger than 0.4(eV) beyond 
the VBM of the wire bandstucture. For this purpose, we compute 50, 100, and 200 energy spectra that correspond to 0.5$\%$, 1.1$\%$, and 2.2$\%$ of the DOF of the mode-space Hamiltonian, 
respectively. Fig. 8(a) shows the corresponding TR and DOS profiles. Here, the CBR solution not only become closer to the reference result with more spectra considered, but also demonstrate 
fairly excellent accuracy near the VBM of the wire bandstructure. The CTR and CDOS profiles provided in Fig. 8(b) further support the preciseness of the CBR solutions near the VBM. The cumulative 
profiles also support that the CBR solution covers a relatively wide range of energy, such that 50 energy spectra are already enough to cover $\sim$0.4(eV) below the VBM quite well. We note that 
the solution obtained with 200 spectra almost replicates the reference result in the entire range of energy that is considered for the simulation ($\sim$0.8(eV) below the VBM).

\begin{figure}[t]
\centering
\includegraphics[width=\columnwidth]{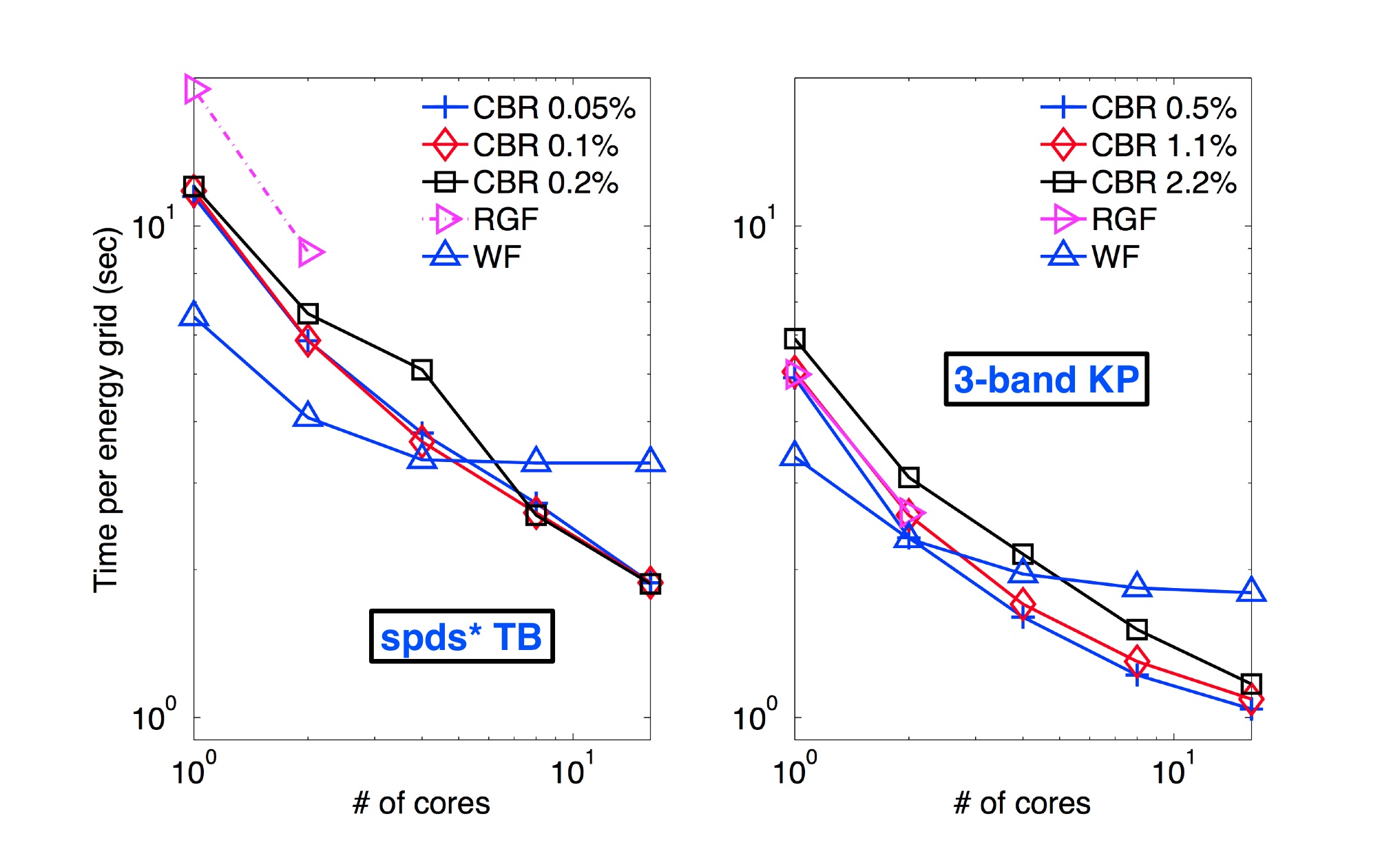}
\caption{
	Speed and scalability of the multi-band CBR method: For the example multi-band systems of TB Si:P RTD and KP 
	Si nanowire FET, we measure the time required to calculate the TR and DOS per single energy point. Scalability of 
	the calculation time is also measured to examine the numerical practicality of the method on HPC clusters.
	}
\end{figure}

\begin{table}[h]
\caption{
	The time required to evaluate the TR and DOS per single energy point in a serial mode, for the RTD and 
	nanowire FET considered as simulation examples.
	} 
\begin{ruledtabular}
\begin{tabular}{llll}
Approaches (TB) & time (s) &\text{ }Approaches (KP) & time (s)\\
\hline
CBR 0.05($\%$) & 11.5 &\text{ }CBR 0.5($\%$) & 4.9\\
CBR 0.1($\%$) & 11.8 &\text{ }CBR 1.1($\%$) & 5.1\\
CBR 0.2($\%$) & 12.0 &\text{ }CBR 2.2($\%$) & 5.9\\
RGF & 19.0 &\text{ }RGF & 5.0\\
WF & 6.5 &\text{ }WF & 3.4
\end{tabular}
\end{ruledtabular}
\end{table}

\emph{Speed and scalability on HPCs}: So far, we have discussed the practicality of the multi-band CBR method focusing on the accuracy of the solutions for two-contact, ballistic-transport problems. 
Another important criterion to determine the numerical utility should be the speed of calculations. We therefore measure the time needed to evaluate the TR and DOS per single energy point for the 
TB Si:P RTD and the KP Si nanowire FET represented that are utilized as simulation examples. To examine the practicality of the multi-band CBR method on HPC clusters, we also benchmark the 
scalability of the simulation time on the $Coates$ cluster under the support of the Rosen Center for Advanced Computing (RCAC) at Purdue University. The CBR, RGF, and WF methods are parallelized 
with MPI/C++, the MUltifrontal Massively Parallel sparse direct linear Solver (MUMPS) \cite{MUMPS}, and a self-developed eigensolver based on the shift-and-invert Arnoldi algorithm \cite{ARNOLDI}. 
All the measurements are performed on a 64-bit, 8-core HP Proliant DL585 G5 system of 16GB SDRAM and 10-gigabit ethernet local to each node.

Table I summarizes the wall-times measured for various methods in a serial mode. Generally, the simulation of the KP Si nanowire FET needs less computing loads, such that the wall-times are 
reduced by a factor of two with respect to the computing time taken for the TB Si:P RTD. This is because the KP approach can represent the electronic structure with the mode-space approach 
such that the Hamiltonian matrix has a smaller DOF (9,000), compared to the one used to describe the TB Si:P RTD (18,720). 

Compared to the RGF algorithm in a serial mode, the CBR method demonstrates a comparable (KP), or better (TB) performance. Since a single slab of the KP Si nanowire is represented with a 
block matrix $H_B$ (Fig. 1) of 120 DOF, the matrix inversion is not a critical problem any more in the RGF algorithm such that the CBR method doesn't necessarily show better performances than 
the RGF algorithm. The TB example device, however, needs a $H_B$ of 720 DOF to represent a single slab (a total of 26 slabs) so the burden for matrix inversions become bigger compared to the 
KP example. As a result, the CBR method generally shows better performances. The CBR method, however, doesn't beat the WF method in both the TB and KP case since, in a serial mode, the CBR 
method consumes time to allocate a huge memory space that is needed to store ``full'' complex matrices via vector-products (Eqs. (5)).

The strength of the CBR method emerges in a $parallel$ mode (on multiple CPUs), where the vector-products are performed via MPI-communication among distributed systems and each node thus 
saves only a fraction of the full matrix. The scalability of the various methods is compared up to a total of 16 CPUs in Fig. 9. The common RGF calculation can be effectively parallelized only up to a 
factor of two, due to its recursive nature \cite{LAKE}, and the scalability of the WF method becomes worse in many CPUs because it uses a direct-solver-based LU factorization to solve the linear 
system. As a result, the CBR method starts to show the best speed when more than 8 CPUs are used.

\section{Conclusion}

In this work, we discuss numerical utilities of the CBR method in simulating ballistic transport of multi-band systems described by the the atomic 10-band ${sp^3d^5s^*}$ TB and 3-band KP approach. 
Although the original CBR method developed for single-band EMA systems achieves an excellent numerical efficiency by approximating solutions of open systems, we show that the same approach 
can't be used to approximate TB systems as the inter-slab coupling matrix becomes singular. We therefore develop an alternate method to approximate open system solutions. Focusing on a proof of 
principles on small systems, we validate the idea by comparing the TR and DOS profile to the reference result obtained by the RGF algorithm, where the alternative also works well with the KP approach.

Since the major numerical issue in the CBR method is to solve a normal eigenvalue problem, the numerical practicality of the method becomes better as the transport can be solved with a less number
of energy spectra. Generally, the practicality would be thus limited in multi-band systems, since multi-band approaches need a larger number of spectra to cover a certain range of energy than the single
band EMA does. We, however, claim that the RTDs could be one category of TB devices, for which the multi-band CBR method becomes particularly practical in simulating transport, and the numerical
utility can be even extended to FETs when the CBR method is coupled to the KP band model. To support this argument, we simulate the electron resonance tunneling in a 3-D TB RTD, which is basically 
a Si nanowire but has a single phosphorous donor in the channel center, and the hole-transport of a 3-D KP Si nanowire FET. We examine numerical practicalities of the multi-band CBR method in terms 
of the accuracy and speed, with respect to the reference results obtained by the RGF and WF algorithm, and observe that the CBR method gives fairly accurate TR and DOS profile near band edges of  
contact bandstructures.

In terms of the speed in a serial mode, the strength of the CBR method over the RGF algorithm depends on the size of the Hamiltonian such that the CBR shows a better performance than the RGF 
as a larger block-matrix is required to represent the unit-slab of devices. But, the speed of the WF method is still better than the CBR method as the CBR method consumes time to store a full complex 
matrix during the process of calculations. In a parallel mode, however, the CBR method starts to beat both the RGF and WF algorithm since the full matrix can be stored into multiple clusters in a distributive 
manner, while the scalability of both the RGF and WF algorithm are limited due to the nature of recursive and direct-solver-based calculation, respectively.

\section*{ACKNOWLEDGEMENTS}

H. Ryu, H.-H. Park and G. Klimeck acknowledge the financial support from the National Science Foundation (NSF) under the contract No. 0701612 and the Semiconductor Research Corporation. 
M. Shin acknowledges the financial support from Basic Science Research Program through the National Research Foundation of Republic of Korea, funded by the Ministry of Education, Science 
and Technology under the contract No. 2010-0012452. Authors acknowledge the extensive use of computing resources in the Rosen Center for Advanced Computing at Purdue University, and 
NSF-supported computing resources on nanoHUB.org. 

%\bibliography{CBR_JAP_Final_HoonRyu}% Produces the bibliography via BibTeX.

\begin{thebibliography}{100}
\bibitem{NOTE01}
In principle, the 4-band KP model uses a total of four bases to model direct bandgap materials such as GaAs and InAs, where 
one basis is used to model the conduction band and the remaining three bases are used to model the valence band. The valence 
band of indirect bandgap materials such Si, however, can be still modeled with three bases if VBM is at the $\Gamma$ point (Ref. [11]).
\bibitem{NOTE03}
Assuming that the source contact is grounded, the Fermi-window at $V_{DS}$ = V, becomes [$E_{FD}$-$3kT$, $E_{FS}$+$3kT$] 
= [$E_{F}$-$qV$-$3kT$, $E_{F}$+$3kT$], where $q$ is the single electron charge and $E_{F}$ is the Fermi-level of the system in 
equilibrium. The maximum and minimum of the window are determined at the source and drain side, respectively.
\end{thebibliography}

\end{document}